\documentclass[prb,twocolumn,showpacs,superscriptaddress,reprint,aps]{revtex4-1}

\usepackage{amsmath, amsfonts, amssymb}
\usepackage{enumerate}
\usepackage{graphicx}
\usepackage{color}
\usepackage{hyperref}
\usepackage{paralist}

\newcommand{\eref}[1]{Eq.~(\ref{#1})}
\newcommand{\fref}[1]{Fig.~\ref{#1}}
\newcommand{\sref}[1]{Sec.~\ref{#1}}

\newcommand{\up}{\uparrow}
\newcommand{\dw}{\downarrow}
\newcommand{\e}{\mathrm{e}}

\newcommand{\si}{\hat{\sigma}_{0}}

\newcommand{\sy}{\hat{\sigma}_{2}}
\newcommand{\sz}{\hat{\sigma}_{3}}

\newcommand{\tx}{\hat{\tau}_{1}}
\newcommand{\ty}{\hat{\tau}_{2}}

\DeclareMathOperator{\real}{Re}
\DeclareMathOperator{\imag}{Im}
\DeclareMathOperator{\sgn}{sign}

\begin{document}


\title{Transport signatures of superconducting hybrids with mixed singlet and chiral triplet states}

\newcommand{\wurzburg}{Institute for Theoretical Physics and Astrophysics,
University of W\"{u}rzburg, D-97074 W\"{u}rzburg, Germany}
\newcommand{\nagoya}{Department of Applied Physics, Nagoya University, Nagoya, 464-8603, Japan}
\newcommand{\tokyo}{Department of Applied Physics, University of Tokyo, Tokyo 113-8656, Japan}
\newcommand{\riken}{RIKEN Center for Emergent Matter Science (CEMS), Wako 351-0198, Japan}

\author{Pablo Burset}
\affiliation{\wurzburg}

\author{Felix Keidel}
\affiliation{\wurzburg}

\author{Yukio Tanaka}
\affiliation{\nagoya}

\author{Naoto Nagaosa}
\affiliation{\tokyo}
\affiliation{\riken}

\author{Bj\"orn Trauzettel}
\affiliation{\wurzburg}

\date{\today}


\begin{abstract}
We propose a model for a superconductor where both spin-singlet and chiral triplet pairing amplitudes can coexist. By solving the Bogoliubov-de Gennes equations with a general pair potential that accounts for both spin states we study experimental signatures of normal metal and superconductor hybrids. 
The interplay between the spin-singlet and triplet correlations manifests in the appearance of two {\it effective} gaps. 
When the amplitude of the spin-triplet component is stronger than that of the spin-singlet, a topological phase transition into a non-trivial regime occurs. As a result, the normal metal-superconductor conductance evolves from a conventional gap profile onto an unconventional zero-bias peak. 
Additionally, in the topologically non-trivial phase, Andreev bound states formed at Josephson junctions present zero-energy modes; the number of those modes depends on the relative chirality of the junction. 
Finally, we present results for the current-phase relation and the temperature dependence of the Josephson critical current within both topological phases for several system parameters. 
\end{abstract}




\maketitle


\section{Introduction\label{sec:intro}}
The symmetry of a Cooper pair is traditionally classified into spin-singlet with orbital even-parity and spin-triplet with odd-parity\cite{Sigrist_RMP}. This classification is valid when the wave function of the Cooper-state can be decomposed into orbital and spin parts. New systems with broken inversion symmetry have been discovered where this classification no longer holds. The Cooper pair in these systems is, therefore, a mixture of singlet and triplet spin states. Such systems include noncentrosymmetric superconductors (NCS) and surface states of topological insulators (TI) in electrical contact with $s$-wave superconductors. 

NCS are superconductors with broken inversion symmetry\cite{Tanaka_JPSJ}. In these materials, the reduced symmetry of the crystal structure, which lacks an inversion center, allows for a robust asymmetric spin-orbit interaction; therefore, the superconducting pair potential mixes singlet and triplet states\cite{Frigeri_2004,*Frigeri_2004b}. 
The relative amplitude between the spin-singlet component of the pair potential $\Delta_s$ and that of the spin-triplet $\Delta_p$ becomes crucial to determine the properties of the NCS\cite{Tanaka_2010,Asano_2011,Annunziata_2012}. 
The surface of a NCS with a mixed singlet and chiral triplet has been predicted to lead to spin-polarized currents\cite{Vorontsov_2008,Tanaka_2009,Chiken_2010}. Furthermore, a two dimensional time-reversal symmetric NCS is expected to host an even number of Majorana fermions\cite{Sato_2009,Santos_2010,Schnyder_2012}. The family of NCS is rapidly increasing and the exact pairing potentials describing many of these materials remains unknown\cite{[{An updated list of recently discovered NCS can be found in }] [{}]Bauer}.

The possibility to induce a triplet state using a conventional superconductor has recently triggered an intense research activity. The most common approach requires conventional s-wave superconductors in proximity with 2D materials with strong spin-orbit coupling\cite{Fujimoto_2008,Sau_2010,Potter_2010,Potter_2011}. 
The interest in these systems dwells in the possibility to control the spin-orbit coupling and, hence, the induced pair potential by means of external magnetic fields. Up to now, the main research line has been focused on engineering an effective spinless $p+ip$ superconductor, which is expected to host topologically protected zero-energy Majorana bound states\cite{Read_2000,*Fu_2008,*Lutchyn_2010,*Alicea_2010,*Oreg_2010,*Duckheim_2011,*Chung_2011,*Ueno_2013}. 
Topologically protected zero-energy boundary modes have been also predicted in NCS\cite{Brydon_2011,Schnyder_2012}. Evidently, there is a strong relationship between superconductivity on the surface of a three-dimensional topological insulator and two-dimensional NCS\cite{Santos_2010}. 

The interplay between the isotropic singlet and the anisotropic triplet spin states is especially relevant near a surface or an interface\cite{Vorontsov_2008}. At the boundary of a superconductor, Andreev reflection opens the possibility for particle-hole coherent conversion. 
These Andreev states manifest in the tunneling spectroscopy of normal metal-superconductor junctions (NS junctions). 
A zero bias conductance peak characterizes the junction between a metal and an unconventional superconductor when the triplet part of the pairing dominates\cite{Annunziata_2012,Tanaka_1995,Kashiwaya_1995,*Kashiwaya_1996}. 
Moreover, Andreev bound states (ABS) are formed at the interface between two superconductors (SNS junctions) notably affecting the Josephson current through the junction. The Josephson current in a junction between superconductors with dominant triplet pairing has been predicted to be carried by single electrons instead of Cooper pairs\cite{Yakovenko_2004}. Therefore, the transport properties of hybrid contacts provide a useful technique for the study of the pairing state. 

Here, we assume that the pair potential at the superconductor is a mixture of spin-singlet isotropic $s$-wave and spin-triplet chiral $p$-wave with out-of-plane orientation. 
Within this assumption, we study transport signatures of both NS and SNS junctions. 
The former are revealed in the tunnel conductance of the NS junction and the latter in the Josephson current. 
The mixing manifests as the appearance of two gaps that can be detected via NS spectroscopy. The ABS at Josephson junctions, on the other hand, are greatly affected by the mixing, inducing a spin asymmetry in the current. As a consequence, the Josephson current-phase relation becomes unconventional at low temperatures. 
Transport signatures in both NS and SNS junctions depend on the degree of mixing of the pair potential, controlled by the amplitudes of each spin-state $\Delta_s$ and $\Delta_p$. We show that the case $\Delta_s=\Delta_p$ is a quantum critical point that distinguishes the topologically trivial phase with $\Delta_s>\Delta_p$ from the non-trivial phase with $\Delta_s<\Delta_p$. Transport signatures strongly depend on this quantum phase transition. 

This article is organized as follows. 
In \sref{sec:model}, we describe the pairing potential for a mixture of singlet and out-of-plane triplet spin states and explain how the Bogoliubov-de Gennes (BdG) equations in spin and Nambu (electron-hole) spaces decouple for this particular choice. 
We solve the BdG equations for a normal metal-superconductor junction in \sref{sec:NS} and explore the impact of the mixture on the conductance of this system. 
In \sref{sec:SNS}, we find the conditions for the formation of ABS in a Josephson junction and their contribution to the supercurrent. We describe the temperature dependence of the Josephson current for several values of the mixing and barrier strength. 
Finally, we conclude with a summary of our results in \sref{sec:conc}. 


\section{Model\label{sec:model}}
\subsection{General considerations}
We work in Nambu (electron-hole) space with basis $\Psi(\mathbf{k})=\left[ u_{\up}(\mathbf{k}), u_{\dw}(\mathbf{k}), v_{\up}(\mathbf{k}), v_{\dw}(\mathbf{k}) \right]^{T}$, where $u_{\sigma}(\mathbf{k})$ and $v_{\sigma}(\mathbf{k})$ are, respectively, the electron and hole-like components with spin $\sigma=\up,\dw$, and $\mathbf{k}$ the wave vector. 
In momentum space, the low-energy excitations of a superconductor are described by the Hamiltonian
\begin{equation}
H(\mathbf{k})=
\left(\! \begin{array}{cc}
        \left[\epsilon(\mathbf{k}) - \mu\right]\si & \hat{\Delta}(\mathbf{k}) \\
	\hat{\Delta}^{\dagger}(\mathbf{k}) & \left[\mu-\epsilon(-\mathbf{k})\right]\si
       \end{array}
\!\right) 
 \,\, ,
\label{eq:BdG}
\end{equation}
where $\epsilon(\mathbf{k})$ is the band energy measured from the chemical potential $\mu$,
$\hat{\dots}$ denotes $2\times2$ matrices and $\si$ is the unit matrix in spin space. 
For a mixture of spin singlet and triplet states, the pairing potential adopts the general form\cite{Frigeri_2004,*Frigeri_2004b,Sigrist_RMP} $\hat{\Delta}(\mathbf{k})=i \left[ \Delta_s(\mathbf{k}) \si + \sum_{j=1}^{3} d_j(\mathbf{k})\hat{\sigma}_j\right]\sy\e^{i\phi}$, with Pauli matrices $\hat{\sigma}_{1,2,3}$ acting on spin space and $\phi$ the superconducting phase. 
The singlet pairing field $\Delta_s(\mathbf{k})$ is an even function of the wave vector. To represent the conventional s-wave superconductivity, we assume the pairing potential to be independent of the wave vector and, thus, $\Delta_s(\mathbf{k})=\Delta_s$ with $\Delta_s$ constant and real. 
On the other hand, the triplet pairing is parametrized\cite{Balian_1963} by an odd vector function $\mathbf{d}(\mathbf{k})=-\mathbf{d}(-\mathbf{k})$. 

In our work, we study a combination of singlet and triplet spin states that allows us to decouple the different spin channels of $H(\mathbf{k})$. Since the singlet state only affects the $\up\dw$ and $\dw\up$ channels, 
we consider, for simplicity, a chiral triplet state of the form $\mathbf{d}(\mathbf{k})=\Delta_p\frac{k_x+i\chi k_y}{|\mathbf{k}|}\mathbf{\hat{z}} =\Delta_p\e^{i\chi\theta}\mathbf{\hat{z}}$ with $\Delta_p\ge0$ the amplitude of the pairing potential and where $\chi=\pm$ labels the opposite chiralities, i.e., the orientation of the angular momentum of the Cooper pairs. 
Consequently, the pairing matrix is 
\begin{equation}
 \hat{\Delta}(\mathbf{k})=i \left[ \Delta_s \si + \Delta_p\e^{i\chi\theta}\sz\right]\sy\e^{i\phi} \,\, ,
\label{eq:pairing}
\end{equation}
which is purely off-diagonal. 
The resulting band dispersion becomes
\begin{equation}
 E_{1,2}(\mathbf{k}) = \sqrt{\epsilon^2(\mathbf{k}) + \Delta_s^2 + \Delta_p^2 \pm 2 \Delta_s\Delta_p\cos\theta} \,\, .
\label{eq:energy}
\end{equation}
As a consequence, \eref{eq:BdG} is decoupled into two spin channels $\up\dw$ and $\dw\up$ with different energies $E_{1}(\mathbf{k})$ and $E_{2}(\mathbf{k})$, respectively. Notice that the change $\mathbf{k}\rightarrow-\mathbf{k}$ exchanges the energy spectra between spin channels due to $\mathbf{d}(\mathbf{k})=-\mathbf{d}(-\mathbf{k})$. From our point of view, this particular choice of the pairing potential is the simplest option that captures the essential and non-trivial physics of mixed pairings at NS and SNS interfaces. 

Interestingly, the interplay between s- and p-wave pairing yields different energy spectra for each spin projection. As a consequence, $\hat{\Delta}(\mathbf{k})$ is not a unitary matrix, i.e., $\hat{\Delta}\hat{\Delta}^{\dagger}=(\Delta_s^2+\Delta_p^2)\si + 2\Delta_s\Delta_p\cos\theta\sz$, and the electronic excitations are affected by two complex pair potentials 
\begin{equation}
 \Delta_{1,2}(\theta)= \left(\Delta_s \pm \Delta_p \e^{i\chi\theta}\right)\e^{i\phi} = |\Delta_{1,2}(\theta)|\e^{i\beta_{1,2}(\theta)}\e^{i\phi} \,\,\, ,
\label{eq:gap-functions}
\end{equation}
with $|\Delta_{1,2}(\theta)|^2=\Delta_s^2+\Delta_p^2 \pm 2\Delta_s\Delta_p\cos\theta$ and \begin{equation}
 \beta_{1,2}(\theta)=\tan^{-1}\frac{\Delta_p\sin\theta}{\Delta_s\pm\Delta_p\cos\theta} \,\, .
\end{equation} 
One can immediately see that $\exp[i\beta_2(\theta)]$ changes sign if $\Delta_s<\Delta_p\cos\theta$. For $\Delta_s=\Delta_p\cos\theta$, the pair potential vanishes for one of the spin channels. 

\begin{figure}
	\includegraphics[width=\columnwidth]{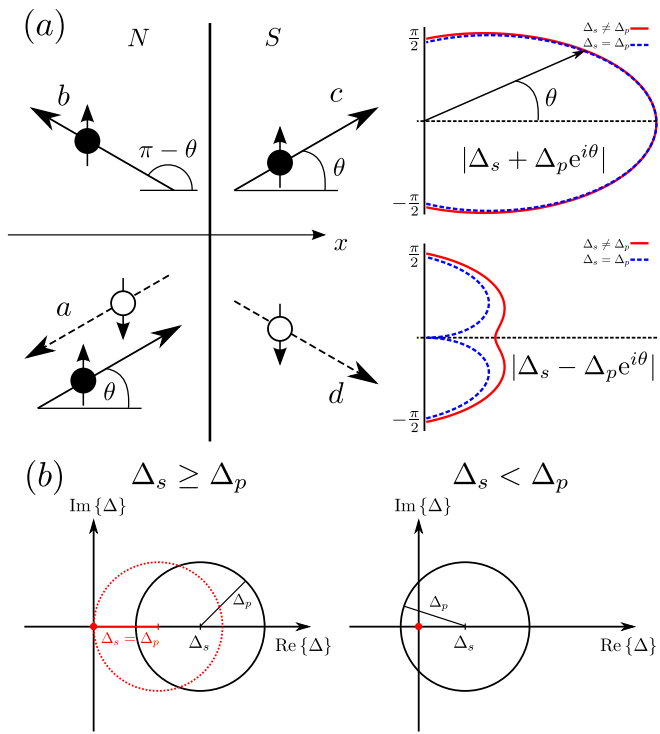}
	\caption{\label{fig:pairing}
	(a) Sketch of the reflection and transmission processes for an incoming spin up electron from the normal metal ($N$). The processes a, b, c, and d denote, respectively, the Andreev reflected hole, the normal reflected electron, the electron-like quasiparticle transmitted into the superconductor ($S$), and the transmitted hole-like quasiparticle. The solid (dashed) arrows represent the velocities of the electrons (holes). 
	A polar plot of the pair potential, with the angle measured with respect to the $k_x$-direction, is shown for the electron- and hole-like excitations of processes c and d. The solid red lines denote an arbitrary situation with $\Delta_s\neq\Delta_p$, while the blue dashed lines correspond to the quantum critical point with $\Delta_s=\Delta_p$. 
	(b) Sketch of the pair potential in the complex plane featuring examples for the three interesting regimes $\Delta_s>\Delta_p$, $\Delta_s=\Delta_p$, and $\Delta_s<\Delta_p$. 
	}
\end{figure}

The Hamiltonian $H(\mathbf{k})$ defined in \eref{eq:BdG} satisfies particle-hole symmetry if $PH^{T}(\mathbf{k})P^{\dagger} = -H(-\mathbf{k})$ with $P$ an unitary operator. There are two possible choices for the particle-hole operator, namely, $P_1 = \tx$ with $P_1P_1^{*}=1$ and $P_2 = \ty \sz$ with $P_2P_2^{*}=-1$ ($\hat{\tau}_{1,2,3}$ are Pauli matrices acting in Nambu space). 
On the other hand, it only satisfies time-reversal symmetry either for $\theta=n\pi$, with $n=0,1,\dots$, or for $\theta=n\pi/2$. In the former case, the $k_y$-component of the chiral spin-triplet state vanishes, while in the latter it is the $k_x$-component. 
Consequently, in the one-dimensional case, $H(\mathbf{k})$ can be classified either in class C, class D and class DIII, according to Cartan-Altland-Zirnbauer\cite{Schnyder_2008,*Budich_2013}, depending on the choice of $\theta$. For the case with $\theta=0$, the Hamiltonian belongs to the non-trivial DIII symmetry class if $\Delta_p>\Delta_s$\cite{Budich_2013b}. 
The transport results presented in the next sections correspond to a two-dimensional system where no specific choice of $\theta$ can be realized. Therefore, $H(\mathbf{k})$ belongs to the overlapping regime between class C or class D. 
To distinguish between these two classes, we compare our transport results in the next section with those of a chiral $d$-wave superconductor, which belongs to class C\cite{[{For more details about junctions with chiral $d$-wave superconductor we refer the reader to }] [{}]Kashiwaya_2014}. 
The transition between trivial and non-trivial topological phases is controlled by the amplitudes $\Delta_s$ and $\Delta_p$. We show in \fref{fig:pairing}(b) a sketch of \eref{eq:gap-functions} in the complex plane. We illustrate both the trivial ($\Delta_s>\Delta_p$) and non-trivial ($\Delta_s<\Delta_p$) cases, and the quantum critical point ($\Delta_s=\Delta_p$). 


\subsection{Quasi-1D limit}
In the following, we consider the quasi-1D limit where transport takes place in the $x$-direction and the transverse component of the wave vector $k_y$ is conserved. To take into account the change of sign of the triplet state with the wave vector, for a fixed $k_y$, we restrict $\mathbf{k}$ to $k_x\ge0$ and, thus, $\alpha\mathbf{k}=(\alpha k_x,k_y)=(\pm k_x,k_y)$. 
Assuming that the band energy $\epsilon(\mathbf{k})$ is the same for left- and right-movers, the change of sign in the triplet component is accounted for by the transformation $\theta\rightarrow\pi-\theta$. We thus define $\theta_{+}=\theta$ for right-movers and $\theta_{-}=\pi-\theta$ for left-movers (see details in \fref{fig:pairing}). 
Since the pairing matrix is anti-diagonal in spin space, we can decouple the two independent spin channels of the Hamiltonian of \eref{eq:BdG} and treat them separately. For each case, we write the $2\times2$ BdG equations
\begin{equation}
\left(\!\! \begin{array}{cc}
        \epsilon(\alpha\mathbf{k}) - \mu & s_{\sigma}\Delta_{\sigma}(\theta_{\alpha}) \e^{i\phi} \\
	s_{\sigma}\Delta^{*}_{\sigma}(\theta_{\alpha}) \e^{-i\phi} & \mu-\epsilon(-\alpha\mathbf{k})
       \end{array}
\!\!\right) \!\!
\left(\!\! \begin{array}{c}
        u_{\sigma}(\theta_{\alpha}) \\ v_{\sigma}(\theta_{\alpha})
       \end{array}
\!\!\right) 
\!\!=\!\! E \!\!
\left(\!\! \begin{array}{c}
        u_{\sigma}(\theta_{\alpha}) \\ v_{\sigma}(\theta_{\alpha})
       \end{array}
\!\!\right) 
\label{eq:BdG_red}
\end{equation}
where $E\geq0$ is the excitation energy, $\alpha=\pm$ for right and left movers, respectively, $\sigma=1,2$ labels the different spin channels and $s_{\sigma}=(-1)^{\sigma-1}$. 
To simplify the following analysis of the pairing potential, we have explicitly written the dependence on the phase factor $\phi$. 

The energy spectrum of \eref{eq:BdG_red} is the same as the one given in \eref{eq:energy}, but now the change in sign is determined by $\alpha$. 
The pairing potentials for each spin channel, given in \eref{eq:gap-functions}, are intimately related to both spin and direction of motion, since $\mathbf{d}(\mathbf{k})$ is an odd function of the wave vector. For example, a right-mover with spin up (down) feels a pairing potential $\Delta_1(\theta_{+})=\Delta_s+\Delta_p\e^{i\chi\theta}$ [$-\Delta_{2}(\theta_{+})=-(\Delta_s-\Delta_p\e^{i\chi\theta})$]; therefore, if it is reflected without spin change it feels a potential $\Delta_1(\theta_{-})=\Delta_s-\Delta_p\e^{-i\chi\theta}$ [$-\Delta_{2}(\theta_{-})=-(\Delta_s+\Delta_p\e^{-i\chi\theta})$]. As a result, the gap amplitude $|\Delta_{1,2}(\theta)|^2$ can be different depending on the direction of motion, as it is shown in the plots of \fref{fig:pairing}(a). This asymmetry reaches a maximum when $\Delta_s=\Delta_p\cos\theta$, where the gap amplitude can even be zero [blue dashed lines in \fref{fig:pairing}(a)]. 

The solutions of \eref{eq:BdG_red} can be described in terms of the amplitudes
\begin{subequations}
\begin{align}
 u_{\sigma}(\theta_{\alpha})= & \frac{1}{\sqrt{2}}\left( 1 + \frac{\sqrt{E^2-|\Delta_{\sigma}(\theta_{\alpha})|^2}}{E} \right)^{1/2} \,\, ,\\
 v_{\sigma}(\theta_{\alpha})= & \frac{1}{\sqrt{2}}\left( 1 - \frac{\sqrt{E^2-|\Delta_{\sigma}(\theta_{\alpha})|^2}}{E} \right)^{1/2} \,\, .
\end{align}
\label{eq:solutions}
\end{subequations}

\section{Normal metal-superconductor junction\label{sec:NS}}
We now apply these results to the conductance of a normal-superconductor junction. Following the formalism introduced by Blonder {\it et al.}\cite{BTK}, we study a one-dimensional metal-insulator-superconductor system. The solution of this 1D model can be extended to higher dimensions provided that there is translational invariance in the directions perpendicular to the electron motion. We assume that the $x$-axis lies in this direction and that the interface is at $x=0$. We model the scattering at the interface using a delta-function potential $V_0(x)=(\hbar^2 k_F/m)  Z\delta(x)$ with $k_F$ the Fermi wave vector, $m$ the electron mass, and $Z$ the dimensionless barrier strength. The normal state metal with $\hat{\Delta}=0$ occupies the $x<0$ region. The pair potential of the superconductor on $x>0$ is the mixture of singlet and chiral triplet spin states introduced in \eref{eq:pairing}. 
When we expand below this formalism to higher dimensions, $\theta$ represents the angle of incidence. 
The different chiralities are connected via a change of sign in $\theta$, so we omit the label $\chi$ in this section. 

We consider electronic excitations near the Fermi surface with electron dispersion relation $\epsilon(\mathbf{k})=(\hbar^2/2m)k_x^2+V_0(x)$. Under the Andreev approximation, which amounts to neglecting terms of order $\Delta_0/\mu$, there is no wave vector mismatch between the normal and the superconducting regions, i.e., $k\equiv k_{N}^{e,h}=k_{S}^{e,h}=k_F$, with $k_F$ the Fermi wave vector. 

The scattering processes resulting from an electron incident on the interface from the normal state region are: 
\begin{inparaenum}[(\itshape a\upshape)]
\item an Andreev reflected hole; 
\item a normal reflected electron; 
\item an electron-like quasiparticle transmitted to the superconductor; and
\item a hole-like transmitted quasiparticle.
\end{inparaenum} 
These processes are sketched in \fref{fig:pairing}(a). 
The reflection amplitudes are obtained solving \eref{eq:BdG_red} with the boundary conditions
\begin{equation}
 \Psi^N=\Psi^S \,\, , \,\, \partial_x\Psi^S-\partial_x\Psi^N=kZ\Psi^N(0) \,\,\, 
\label{eq:bbcc}
\end{equation}
with $\Psi^N$ and $\Psi^S$ the wave function at the normal and superconducting sides of the interface. Namely,
\begin{subequations}
\begin{align}
 \Psi_{\sigma}^N(x) ={}& \e^{ikx}\!\! \left[ \left(\!\!\begin{array}{c} 1 \\ 0 \end{array} \!\!\right) \!+\! 
 a_{\sigma}(E) \left(\!\!\begin{array}{c} 0 \\ 1 \end{array} \!\!\right) \right] \!\!+\! 
 b_{\sigma}(E) \e^{-ikx}\!\! \left(\!\!\begin{array}{c} 1 \\ 0 \end{array} \!\!\right) \\
 \Psi_{\sigma}^S(x) ={}& 
 c_{\sigma}(E) \e^{ikx}\!\! \left(\!\!\! \begin{array}{c} u_{\sigma}(\theta_{+}) \\ \eta^{*}_{\sigma}(\theta_{+})v_{\sigma}(\theta_{+}) \end{array} \!\!\!\right) \nonumber \\
 &+d_{\sigma}(E) \e^{-ikx}\!\! \left(\!\!\! \begin{array}{c} \eta_{\sigma}(\theta_{-})v_{\sigma}(\theta_{-}) \\ u_{\sigma}(\theta_{-}) \end{array} \!\!\!\right)
\label{eq:wave-functions_NS}
\end{align}
\end{subequations}
where $\eta_{\sigma}(\theta_{\alpha})=s_{\sigma}\Delta_{\sigma}(\theta_{\alpha})/|\Delta_{\sigma}(\theta_{\alpha})|$. 
Using that $\Delta_{1,2}(\theta_{-})=\Delta_{2,1}^{*}(\theta)$, the resulting reflection amplitudes are
\begin{subequations}
\begin{align}
 a_{\sigma=1,2}(E,\theta) &= \frac{ 4\eta^{*}_{1,2}(\theta) v_{1,2}(\theta) u_{2,1}(\theta) }{ 4 u_{1}(\theta) u_{2}(\theta) + Z^2 t_{\sigma}(E,\theta) }  \, ,
\\
 b_{\sigma}(E,\theta) &= \frac{ -Z(Z+2i)t_{\sigma}(E,\theta) }{ 4 u_{1}(\theta) u_{2}(\theta) + Z^2 t_{\sigma}(E,\theta) }
 \, , \\
 t_{\sigma}(E,\theta) &= u_{1}(\theta) u_{2}(\theta) - \eta^{*}_{1}(\theta)\eta^{*}_{2}(\theta)v_{1}(\theta) v_{2}(\theta) \, .
\label{eq:amplitudes_NS}
\end{align}
\end{subequations}
On the basis of \eref{eq:solutions}, one immediately obtains that $t_{1}(E,\theta) =t_{2}(E,\theta) $ and, consequently, $b_{1}(E,\theta) =b_{2}(E,\theta) $. This relation does not hold for the Andreev reflection amplitude. However, for $|E|\leq|\Delta_{2}(\theta)|$, one finds that $|a_{1}(E,\theta)|^{2} =|a_{2}(E,\theta)|^{2}$. 

Following Refs.~\onlinecite{Tanaka_1995} and \onlinecite{Kashiwaya_1995,*Kashiwaya_1996}, 
for $E<\Delta_0$ and $Z\neq0$, one obtains perfect Andreev reflection ($|a(E)|^2=1$) provided that $ t_{\sigma}(E,\theta) = 0$. For large $Z$, this condition is equivalent to the formation of a bound state at the surface of a semi-infinite superconductor. Any real solution of this equation is associated with the formation of a sub-gap resonance at the interface\cite{Kashiwaya_1995,*Kashiwaya_1996,Honerkamp_1998}. 
This resonance condition can be compactly written as
\begin{equation}
 \frac{\Delta^{*}_{1}(\theta)}{\Delta_{2}(\theta)} = \frac{E+i\sqrt{|\Delta_{1}(\theta)|^2-E^2}}{E-i\sqrt{|\Delta_{2}(\theta)|^2-E^2}} \,\,\, .
\label{eq:resonance}
\end{equation}
When the triplet component is stronger than the singlet one ($\Delta_p\cos\theta>\Delta_s$), this equation has a real solution of the form $E_0=\Delta_p\sin{\theta}$. Since $t_{1}(E,\theta) =t_{2}(E,\theta) $, the resonance condition is the same for both spin channels. Therefore, for $\theta\neq0$ the reflection amplitudes become asymmetric with respect to the energy, revealing the chiral behavior of the pairing potential\cite{Honerkamp_1998}. 

At zero temperature, the single-mode conductance of the system can be obtained by the superposition of the contribution from each spin channel
\begin{equation}
  G_{NS}(E,\theta)\!=\!\frac{e^2}{h}\sum\limits_{\sigma}\left(1 \!+\! \left|a_{\sigma}(E,\theta)\right|^2 \!-\! \left|b_{\sigma}(E,\theta)\right|^2 \right) .
\label{eq:cond-1D}
\end{equation}

When the barrier at the interface does not mix different modes, this result can easily be generalized to higher dimensions. 
Assuming that the momentum component parallel to the interface is conserved, all wave vectors for a given mode lie in the same plane. Under the Andreev approximation, the angle of incidence of incoming quasiparticles from the normal region is the same as the transmitted excitations into the superconductor (see \fref{fig:pairing}). 
On the other hand, the orientation of the triplet component of the pairing potential is defined relative to the NS interface. 
Therefore, the angle $\theta$ can be associated with the angle of incidence if both are measured with respect to the $k_x$-direction, i.e., $\e^{i\theta}=(k_x+ik_y)/k_F$. Consequently, the contribution from multiple modes is given by the angle average of the single-mode conductance as\cite{BTK,Honerkamp_1998}
\begin{equation}
  \tilde{G}_{NS}(E)=\int\limits_{-\pi/2}^{\pi/2} P(\theta)G_{NS}(E,\theta) \cos\theta \mathrm{d}\theta \,\, .
\label{eq:conductance}
\end{equation}
$P(\theta)$ is the experiment-dependent probability distribution; in what follows, we assume $P(\theta)=1$. The conductance, in the subsequent discussion of the results, is normalized by the normal state conductance $G_0=(2e^2/h)D(\theta)$, with $D(\theta)=4\cos^2\theta/(Z^2+4\cos^2\theta)$ the normal state transmission for a single mode in the quasi-1D limit.

\begin{figure}
	\includegraphics[width=\columnwidth]{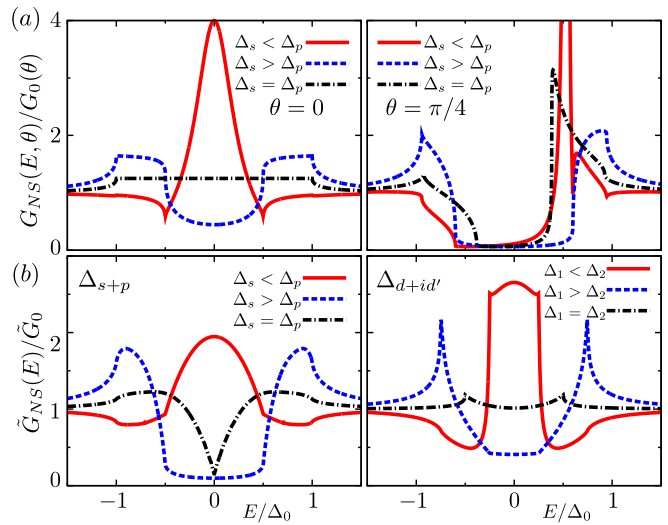}
	\caption{\label{fig:conductance}
	(a) Normalized single-mode conductance for the cases $\Delta_p=1-\Delta_s=0.75$ (red solid line), $\Delta_p=1-\Delta_s=0.25$ (blue dashed line), and $\Delta_p=\Delta_s=0.5$ (black dashed-dot line). $Z=2$ and $\theta=0$ for the left panel; $Z=4$ and $\theta=\pi/4$ for the right one. 
	(b) Angle-averaged conductance for a junction with $Z=4$. Left panel corresponds to the case where the pairing potential mixes singlet and triplet states; the amplitudes $\Delta_s$ and $\Delta_p$ are the same as in (a). For the right panel, the pairing potential is a chiral $d$-wave with $\Delta_2=1-\Delta_1=0.75$ (red solid line), $\Delta_2=1-\Delta_1=0.25$ (blue dashed line), and $\Delta_1=\Delta_2=0.5$ (black dashed-dot solid line). 
	}
\end{figure}

As we show now, the effect of the two gaps in the energy spectrum and the formation of sub-gap resonances can be nicely seen in the conductance. 
We first consider the single mode case with $\theta=0$. The energy spectrum of \eref{eq:energy} then reduces to $E_{1,2}= \sqrt{E^2-\Delta_{1,2}^2} = \sqrt{E^2-\left(\Delta_s \pm \Delta_p\right)^2}$. 
In \fref{fig:conductance}(a), we plot the conductance normalized to the normal state conductance $G_0$, calculated with $Z=2$. We consider three situations depending on the relative values of $\Delta_s$ and $\Delta_p$. 
For $\Delta_s>\Delta_p$ (blue dashed line) the conductance is strongly suppressed for $E<\Delta_{2}$, similarly to the case of a junction with a conventional s-wave superconductor. The case with $\Delta_s <\Delta_p$ (red solid line), however, presents a zero bias conductance peak as it is expected for an unconventional superconducting junction\cite{Tanaka_1995,Kashiwaya_1995,*Kashiwaya_1996,Yamashiro_1997}. The appearance of this peak is associated with the formation of a sub-gap resonance at $E=0$ and the width of the resonance decreases as $Z^{-2}$. 
In the range $\Delta_{2}<E<\Delta_{1}$, Andreev reflection is strongly suppressed for the excitations with energy dispersion $E_{2}$. Incident quasiparticles in this energy branch are no longer affected by the pairing potential and can not be Andreev reflected. The incoming quasiparticles with $E_1$ are still affected by the pairing potential and can be Andreev reflected; therefore, the conductance slowly increases. For $E>\Delta_{1}$ the conductance reduces to the normal state conductance $G_0$. 
Finally, in the case where $\Delta_s=\Delta_p$, one of the energy branches is no longer affected by the pairing potential. For this gapless channel, $a(E)=0$ and the transmission becomes $D=1-|b(E)|^2=4/(Z^2+4)$, which provides a constant contribution to the conductance as in the normal state. For the other channel, both Andreev and normal reflections are constant for $|E|\le|\Delta_1|$. The resulting conductance for both spin channels is plotted as the black dashed-dot line in the left panel of \fref{fig:conductance}(a). 


When the phase factor $\theta$ of the triplet state is finite, the reflection amplitudes for each spin channel become asymmetric with respect to the energy. We show in the right panel of \fref{fig:conductance}(a) the single-mode conductance for $Z=4$ and $\theta=\pi/4$ with the same color scheme as before. The red solid line for $\Delta_s<\Delta_p$ clearly shows a sub-gap resonance at $E=\Delta_p/\sqrt{2}$. By decreasing the difference between $\Delta_s$ and $\Delta_p$, the resonance smoothly merges with the continuum at $E\ge|\Delta_2(\theta)|$. 
The asymmetry with the energy is maintained in the regime $\Delta_s>\Delta_p$ even though there are no sub-gap resonances. 

In the left panel of \fref{fig:conductance}(b), we show the angle averaged conductance for a junction with $Z=4$. The transition from a {\it gapped} conductance profile for $\Delta_s>\Delta_p$ (blue dashed line) into a zero bias conductance peak for $\Delta_s<\Delta_p$ (red solid line) is still reproduced. The situation with $\Delta_s=\Delta_p$ (black dashed-dot line) develops an interesting sub-gap structure where there are no resonances but which is not fully gapped. This corresponds to the quantum critical point where the bulk gap is closed, but the condition for the formation of subgap resonances is not yet fulfilled. 
For comparison, we have included in the left panel of \fref{fig:conductance}(b) the conductance of a junction with a chiral $d$-wave superconductor, which belongs to symmetry class C. The pairing potential for this case adopts the form\cite{Kashiwaya_2014} $\Delta_{d+id'}(\theta_{\pm})=\Delta_1\cos(2\theta_{\pm})+i\Delta_2\sin(2\theta_{\pm})$, which presents a chiral structure similar to the mixing potential that we are using. A transition from a gapped profile into a conductance peak can also be reproduced by changing the amplitudes $\Delta_1$ and $\Delta_2$ [right panel of \fref{fig:conductance}(b)]. It is interesting to note that when $\Delta_1=\Delta_2$, the resulting conductance is almost flat. This is in contrast to the $\Delta_s=\Delta_p$ case, which features a V-shaped zero-energy dip [black dot-dashed lines in \fref{fig:conductance}(b)].

\section{Josephson junction\label{sec:SNS}}
We now consider a junction between two superconductors. 
As it was the case for the NS junction, the junction barrier is located at $x=0$ with one of the superconductors ($L$) located in the region $x<0$ and the other ($R$) at $x>0$. 
The pair potential at each superconductor is a combination of singlet and triplet states, as in \eref{eq:pairing}, and we can thus treat each spin channel separately. 
The pairing potential in \eref{eq:BdG_red} then adopts the form 
\begin{equation}
 \Delta_{1,2}(x,\theta_{\alpha})=\left\{ \begin{array}{cr}
		    (\Delta_s^L\pm\Delta_p^L\e^{i\chi_L\theta_{\alpha}})\e^{i\phi_L}\, , & x<0 \, . \\ 
		    (\Delta_s^R\pm\Delta_p^R\e^{i\chi_R\theta_{\alpha}})\e^{i\phi_R}\, , & x>0 \, .
                            \end{array} \right.
\end{equation}
The electrical current flowing through the junction depends on the phase difference between the two superconductors $\phi=\phi_R-\phi_L$. It also depends on the relative chirality of the superconductors: we can have junctions with parallel chirality ($\chi_L\chi_R=1$) and junctions with opposite chirality ($\chi_L\chi_R=-1$). To distinguish each case it is enough to assume $\chi_L=1$ and $\chi_R=\chi=\pm$. According to \eref{eq:gap-functions}, a change of chirality is equivalent to a change in the sign of $\theta$; therefore, we define $\tilde{\theta}_{+}=\chi\theta$ and $\tilde{\theta}_{-}=\pi-\chi\theta$ for the superconductor R. 

\subsection{Contributions to the Josephson current.}
At each superconductor we have two effective gaps $|\Delta^{L,R}_{1,2}(\theta)|$, with $|\Delta^{L,R}_{1}(\theta)|\geq|\Delta^{L,R}_{2}(\theta)|$. For simplicity, we consider symmetric junctions where the amplitude of the pairing potential is the same on both sides of the junction; $\Delta_s^{L}=\Delta_s^{R}\equiv\Delta_s\ge0$ and $\Delta_p^{L}=\Delta_p^{R}\equiv\Delta_p\ge0$. We can thus define three energy regimes: 
\begin{inparaenum}[(\itshape 1\upshape)]
\item $|E|\le|\Delta_{2}(\theta)|$; 
\item $|\Delta_{2}(\theta)|<|E|\le|\Delta_{1}(\theta)|$; and
\item $|E|>|\Delta_{1}(\theta)|$. 
\end{inparaenum} 
Accordingly, the total Josephson current can be divided into three contributions $I(\phi)=I_1(\phi)+I_2(\phi)+I_3(\phi)$. 

The Josephson currents $I_1(\phi)$ and $I_3(\phi)$ correspond to the contributions from discrete Andreev levels within the gap and excited states from the continuum, respectively. The continuum contribution $I_3(\phi)$ is negligible for the short ballistic junction considered here\cite{Zagoskin}. 
The current carried by each Andreev state is $(e/h)\partial E_{\sigma,n}(\phi)/\partial \phi$, where $E_{\sigma,n}(\phi)$ is the corresponding energy level\cite{Kulik_1975,Beenakker_1991,Bagwell_1992,Zagoskin}. Therefore, we have
\begin{equation}
 I_1(\phi)= \frac{e}{h}\sum\limits_{\sigma,n} \int^{\pi/2}_{-\pi/2} \frac{\partial E_{\sigma,n}(\phi,\theta)}{\partial \phi} f(E_{\sigma,n}) \mathrm{d}\theta\cos\theta \quad ,
\label{eq:current-abs}
\end{equation}
where $\sigma=\up\dw,\dw\up$ labels the spin channel, $n$ the energy level, and $f(E_{\sigma,n})=[1+\exp(E_{\sigma,n}/k_BT)]^{-1}$ is the equilibrium Fermi occupation factor, with $k_B$ the Boltzmann constant and $T$ the temperature. 

However, in the regime where $|\Delta_{2}(\theta)|<|E|\le|\Delta_{1}(\theta)|$, the excitations at the interface between superconductors are either Andreev reflected or transmitted into the superconductor depending on their spin and direction of motion. 
We show in Appendix \ref{sec:app1} that the contribution to the current in this energy range is zero for transparent junctions, i.e., when $Z=0$. That is not the case for junctions with arbitrary barrier strength $Z$. 
For these junctions, this contribution must be taken into account when computing the Josephson current. 
A similar separation of contributions to the Josephson current is reached in an asymmetric junction where the pair potential is different in each superconductor\cite{Chang_1994}. 

\subsection{Andreev bound states.}
The wave function for each superconductor is a superposition of the solutions of the BdG equations given in \eref{eq:solutions}
\begin{subequations}
\begin{align}
 \Psi^{L}_{\sigma}(x) ={}& 
C_{L}^{+} \left(\!\! \begin{array}{c}
          \eta^{L}_{\sigma}(\theta_{+})v^{L}_{\sigma}(\theta_{+}) \\ u^{L}_{\sigma}(\theta_{+})
\end{array}\!\! \right) \e^{\frac{\Omega^{L}_{\sigma}(\theta_{+})x}{\hbar v_F}} \e^{ikx} 
\nonumber \\
+& C_{L}^{-} \left(\!\! \begin{array}{c}
          u^{L}_{\sigma}(\theta_{-}) \\ \eta^{L*}_{\sigma}(\theta_{-})v^{L}_{\sigma}(\theta_{-})
\end{array}\!\!\right) \e^{\frac{\Omega^{L}_{\sigma}(\theta_{-})x}{\hbar v_F}} \e^{-ikx} \,\,, 
\end{align}
\begin{align}
 \Psi^{R}_{\sigma}(x) ={}& 
C_{R}^{+} \left(\!\! \begin{array}{c}
          u^{R}_{\sigma}(\theta_{+}) \e^{i\phi} \\ \eta^{R*}_{\sigma}(\tilde{\theta}_{+})v^{R}_{\sigma}(\theta_{+})
\end{array}\!\! \right) \e^{-\frac{\Omega^{R}_{\sigma}(\theta_{+})x}{\hbar v_F}} \e^{ikx} 
\nonumber\\
+&
C_{R}^{-} \left(\!\! \begin{array}{c}
          \eta^{R}_{\sigma}(\tilde{\theta}_{-})v^{R}_{\sigma}(\theta_{-}) \e^{i\phi} \\ u^{R}_{\sigma}(\theta_{-})
\end{array}\!\!\right) \e^{-\frac{\Omega^{R}_{\sigma}(\theta_{-})x}{\hbar v_F}} \e^{-ikx} \,\, ,
\end{align}
\label{eq:wave-functions_J}
\end{subequations}
where $\Omega^{L,R}_{\sigma}(\theta_{\alpha})=\sqrt{|\Delta^{L,R}_{\sigma}(\theta_{\alpha})|^2-E^2}$. 
Substituting the wave functions in the boundary conditions of \eref{eq:bbcc} we obtain a system of linear homogeneous equations for the coefficients $C_{L,R}^{\pm}$. This system has a non-trivial solution if the determinant of the associated matrix is zero. 

For symmetric junctions with $\chi=+1$, this condition is reduced to 
\begin{align}
\real \left\{\Delta_{1}\Delta_{2}\right\} ={}& A_{-} + D\left[ \real \left\{\Delta_{1}\Delta_{2}\right\} \right. \nonumber\\
& \left.- A_{+}\cos{\phi}+s_{\sigma}B_{-}\sin{\phi} \right] \, ,
\label{eq:comp-red} 
\end{align}
where we have omitted the dependence on $\theta$ for simplicity and we defined
 \begin{align*}
  A_{\pm} &= E^2 \pm \Omega_{1}(E,\theta)\Omega_{2}(E,\theta) \,\, , \\
  B_{\pm} &= E\left[ \Omega_{1}(E,\theta)\pm\Omega_{2}(E,\theta)\right] \,\, .
 \end{align*}

On the other hand, for $\chi=-1$, we find
\begin{align}
|\Delta_1\Delta_2|^2 - \real \left\{\Delta_{1}\Delta_{2}\right\} A_{-} - \imag \left\{\Delta_{1}\Delta_{2}\right\} B_{+} = 
\nonumber\\
D\left( |\Delta_1\Delta_2|^2 - \real \left\{\Delta_{1}\Delta_{2}^{*}\right\}\left[ 
A_{+}\cos{\phi} - s_{\sigma}B_{-}\sin{\phi} \right] \right. \nonumber\\ 
\left. + \imag \left\{\Delta_{1}\Delta_{2}^{*}\right\}\left[ B_{-}\cos{\phi} + s_{\sigma}A_{+}\sin{\phi} \right] \right) \, .
\label{eq:comp-red-chi} 
\end{align}

The solutions $E_{\sigma}$ of \eref{eq:comp-red} and \eref{eq:comp-red-chi} form the ABS of the junction for each spin projection. 
For the junction with $\Delta_p=0$ and $\Delta_s$ finite, these solutions are the well known ABS for a one-dimensional Josephson junction between $s$-wave superconductors\cite{Yakovenko_2004}
\begin{equation}
 E_{\sigma}^{\pm}(\phi)=\pm\Delta_s\sqrt{ 1-D\sin^2{(\phi/2}) } \,\,\, ,
\label{eq:ABS_ssJ}
\end{equation}
which are spin and angle independent. For a transparent junction ($D=1$) the equation for the bound states reduces to $E_{\sigma}^{\pm}(\phi)=\pm\Delta_p\left| \cos{(\phi/2})\right|$. The positive and negative branches touch at $E=0$ but do not change sign; the energy levels are thus $2\pi$-periodic. 

On the other hand, if $\Delta_s=0$ with a finite $\Delta_p$ and $\theta=0$ the bound states are
\begin{equation}
 E_{\sigma}^{\pm}(\phi)=\pm\Delta_p\sqrt{ D } \cos{(\phi/2})  \,\,\, .
\label{eq:ABS_ppJ}
\end{equation}
The same expression is found for the ABS formed at a $p_x$-$p_x$ junction\cite{Yakovenko_2004} or at the $45^{\circ}/45^{\circ}$ junction between two $d$-wave superconductors\cite{Tanaka_1996}. Independently of the transmission of the junction the energy levels change sign at $\phi=2n\pi$, with $n=0,\pm1,\dots$, and are $4\pi$-periodic. 
Moreover, for $\Delta_s=0$, $\Delta_p\neq0$, and arbitrary $\theta$, the roots of \eref{eq:comp-red-chi} and \eref{eq:comp-red} reproduce the analytical expressions for the ABS of a junction between chiral $p$-wave superconductors [Eqs. (47) and (48), respectively, in Ref.~\onlinecite{Yakovenko_2004}]. Finally, for $D=0$, the roots for both chiralities are given by $E^{\pm}=\pm\Delta_p\sin\theta$, which represent the chiral Andreev surface states at each superconductor. 


\begin{figure}
	\includegraphics[width=\columnwidth]{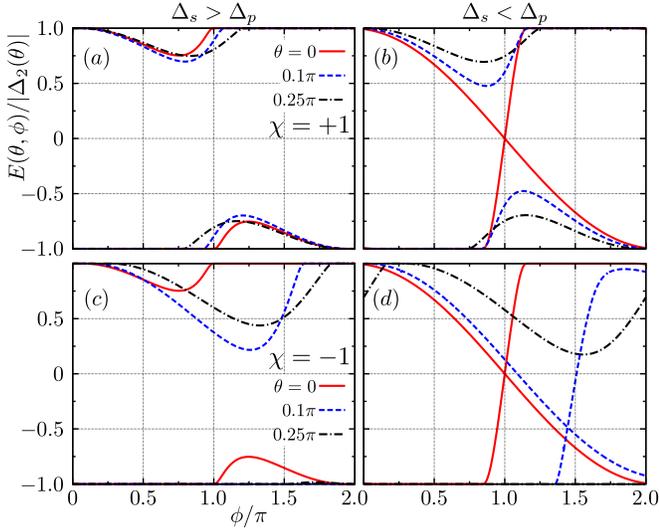}
	\caption{\label{fig:ABS-Z1}
	ABS for symmetric junctions as a function of $\phi$ for severaal values of the angle $\theta$. We show $E^{+}_1(\theta,\phi)$ and $E^{-}_2(\theta,\phi)$ for $\chi=+1$ and $\Delta_s=1-\Delta_p=0.55$ (a); $\chi=+1$ and $\Delta_s=0.45$ (b); $\chi=-1$ and $\Delta_s=0.55$ (c); and, $\chi=-1$ and $\Delta_s=0.45$ (d). For all plots we set $Z=1$. 
	}
\end{figure}


For symmetric transparent ($D=1$) junctions with $\chi=+1$, the solutions of \eref{eq:comp-red} adopt the form 
\begin{equation}
 E_{1,2}^{\pm}(\phi,\theta)=\pm|\Delta_{1,2}(\theta)|\cos(\phi/2) \quad .
\label{eq:ABS-cp1}
\end{equation}
For this special case, the ABS are always zero at $\phi=\pi$. When the spin-states mix, the ABS become different [i.e., $E_1(\theta,\phi)\neq E_2(\theta,\phi)$], with the splitting controlled by the mixing and the angle of incidence. 

For the case with $\chi=-1$, the solutions of \eref{eq:comp-red-chi} when $D=1$ read
\begin{gather}
 E_{1,2}^{\pm}(\phi,\theta)= \nonumber \\ \frac{ \Delta_s^2\sin\phi + \Delta_p^2\sin(\phi\mp2\theta) \pm 2\Delta_s\Delta_p\sin(\phi\mp\theta) }{ 2|\Delta_s\sin\frac{\phi}{2} \pm \Delta_p\sin(\frac{\phi}{2}\mp\theta) | } \quad .
\label{eq:ABS-cm1}
\end{gather}
For $\theta=0$ the ABS reduce to the simple form of \eref{eq:ABS-cp1}. For a finite angle, however, the periodicity of the ABS $E^{\pm}_1$ and $E^{\pm}_2$ is shifted differently. Additionally, as in the previous case, the amplitude of the ABS is also changed by the mixing of spin-states. 

For finite barrier strength ($D\neq1$), the solutions of \eref{eq:comp-red} and \eref{eq:comp-red-chi} are obtained numerically. We show in \fref{fig:ABS-Z1} the ABS $E^{+}_1(\theta,\phi)$ and $E^{-}_2(\theta,\phi)$ as a function of the phase difference $\phi$ for several values of the angle of incidence $\theta$. We normalize the ABS to the value of the bulk gap $|\Delta_2(\theta)|$. When we introduce a finite mixing of spin-states, controlled by the amplitude $\Delta_s=1-\Delta_p$, the ABS immediately become different independently of the rest of parameters. For $\theta=0$ (red solid lines), the ABS fulfill $E_1^{+}(0,\phi)=E_2^{-}(0,-\phi)$. Compared with the transparent case, the ABS develop a gap when $\Delta_s>\Delta_p$, but remain gapless for $\Delta_s<\Delta_p$. For junctions with $\chi=+1$, the gapless mode disappears when $\theta\neq0$ [see \fref{fig:ABS-Z1}(b)]. 
Additionally, for these junctions the ABS are asymmetric with respect to $\theta$ and fulfill $E_1^{+}(\theta,\phi)=E_2^{-}(-\theta,-\phi)$. 
Interestingly, the spectrum of junctions with $\chi=-1$ remains gapless for a wide range of the angle $\theta$ due to the chiral dispersion of the spin-triplet component. 
The number of surface states in isolated chiral superconductors is given by the Chern number\cite{Tanaka_JPSJ,Yoo_2014}. For the chiral triplet cases studied here this number can be $n_{L,R}=\pm1$, depending on the chirality of the superconductor. At the interface between two chiral superconductors, the number of bound states is determined by $|n_L-n_R|$. For $\chi=+1$, we have $|n_L-n_R|=0$ and the only zero-energy solutions are restricted to the values $\theta=0$ and $\phi=\pi$. On the other hand, for $\chi=-1$ we find $|n_L-n_R|=2$. Consequently, we find zero-energy solutions for a wider range of $\theta$. 


\begin{figure}
	\includegraphics[width=\columnwidth]{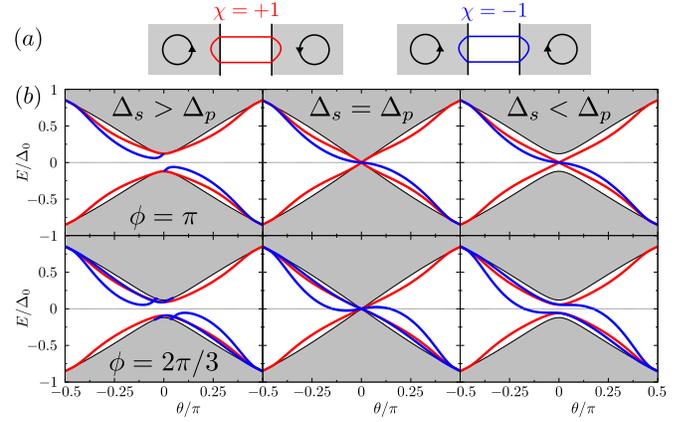}
	\caption{\label{fig:ABS-gap}
	(a) Sketch of the formation of a bound states at junctions with the same (left) and opposite (right) chiralities. (b) ABS $E^{\pm}_{1,2}(\theta,\phi)$ compared to the bulk gap $|\Delta_2(\theta)|$ as a function of the angle of incidence $\theta$. Following the sketch from (a), ABS for junction with $\chi=+1$ ($\chi=-1$) are shown in red (blue) lines. From left to right we show the trivial case with $\Delta_s=0.55$, the quantum critical point $\Delta_s=0.5$ and the non-trivial case with $\Delta_s=0.45$. The top panels correspond to $\phi=\pi$ and the bottom to $\phi=2\pi/3$. For all cases $Z=1$.
	}
\end{figure}


The form of the ABS strongly depend on the mixing of spin-states and on the chirality of the junction. We show in \fref{fig:ABS-gap}(a) a sketch of the formation of ABS at junctions with parallel (left panel) or opposite chirality (right). The mixing of spin-states controls the value of the bulk gap, given by $|\Delta_2(\theta)|$, distinguishing the topologically different regions with $\Delta_s>\Delta_p$ and $\Delta_s<\Delta_p$. The bulk gap closes at the quantum critical point $\Delta_s=\Delta_p$. We show in \fref{fig:ABS-gap}(b) the ABS as a function of $\theta$ for different values of the phase difference $\phi$. 
We normalize the ABS to $\Delta_0\equiv\sqrt{\Delta_s^2+\Delta_p^2}\Delta(T=0)$, where $\Delta(T=0)$ provides the right units. The particular choice of $\Delta(T=0)$ is irrelevant; throughout the paper we use a value comparable to that of $s$-wave superconductors at zero temperature. 
In the topologically trivial case with $\Delta_s>\Delta_p$, junctions with different chiralities display a different dispersion of the ABS, but they never present zero-energy modes. For a fixed value of $\phi$, the ABS are symmetric with respect to $\theta$ for $\chi=+1$ and anti-symmetric when $\chi=-1$. For $\Delta_s=\Delta_p$, the bulk gap closes at $\theta=0$. In this case, junctions with $\chi=-1$ feature zero-energy modes for $\theta\neq0$, where the bulk gap is open. Finally, for the topologically non-trivial case with $\Delta_s<\Delta_p$, there is always a zero energy mode at $\theta=0$ and $\phi=\pi$, independently of the chirality. This gapless dispersion disappears for $\phi\neq\pi$ when $\chi=+1$. However, junctions between superconductors with opposite chirality have at least two zero-energy solutions that become degenerate at $\phi=\pi$. 


\begin{figure}
	\includegraphics[width=\columnwidth]{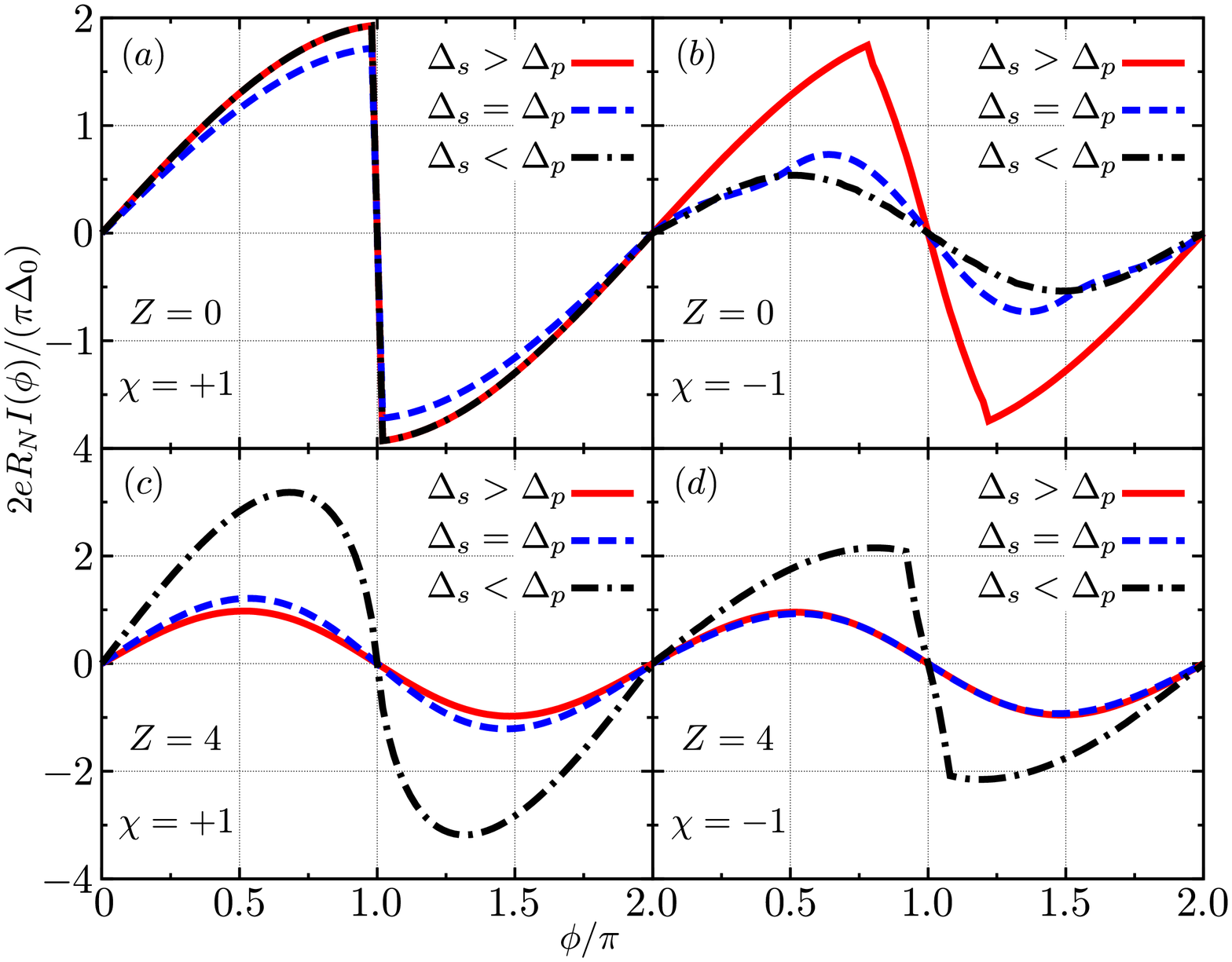}
	\caption{\label{fig:current-phi}
	Josephson current through symmetric junctions for $\Delta_s=1-\Delta_p=0.75$ (red solid lines), $\Delta_s=0.5$ (blue dashed lines), and, $\Delta_s=0.25$ (black dot-dashed lines). For $\chi=+1$, we show the current for junctions with $\chi=+1$ (a) and $\chi=-1$ (b). For $Z=4$, the same cases in (c) and (d), respectively. For all cases, $T/T_c=0.001$. 
	}
\end{figure}


\subsection{Josephson current}
The ABS come in pairs $E^{\pm}_{1,2}(\phi,\theta)$ for each spin channel. Inserting these solutions into \eref{eq:current-abs} we obtain
\begin{align}
 I_1(\phi)={}& \frac{e}{2h}\int^{\pi/2}_{-\pi/2} \mathrm{d}\theta\cos\theta \nonumber \\
  & \times \left[ \partial_{\phi}E_{1}(\phi,\theta)\tanh\left(\frac{E_{1}(\phi,\theta)}{2k_BT} \right) \right. \nonumber \\
 & + \left. \partial_{\phi}E_{2}(\phi,\theta)\tanh\left(\frac{E_{2}(\phi,\theta)}{2k_BT} \right)  \right] \,\, ,
\label{eq:current-abs-D1}
\end{align}
where both spin channels have been taken into account. In \fref{fig:current-phi}(a,b) we show the total current for transparent junctions ($Z=0$). For these junctions, the only contribution to the Josephson current comes from the ABS as \eref{eq:current-abs-D1} (see Appendix \ref{sec:app1}). Following the analysis of the ABS, the current is very different for junctions with parallel or opposite chirality. Transparent junctions with $\chi=+1$ do not show different behavior in the regimes $\Delta_s\lessgtr\Delta_p$. In \fref{fig:current-phi}(a) we show that the case with $\Delta_s=1-\Delta_p=0.75$ is the same as that of $\Delta_s=0.25$ (red solid and black dot-dashed lines, respectively). Both currents are given by the ABS in \eref{eq:ABS-cp1}, where the mixing only affects the amplitude of the ABS. Therefore, the amplitude of the current reaches a minimum for $\Delta_s=\Delta_p$ (blue dashed line). When the current is given by the ABS of \eref{eq:ABS-cp1}, the profile is highly non-sinusoidal at low temperatures. 
The situation is very different when $\chi=-1$, where the ABS are given by \eref{eq:ABS-cm1}. When we consider a small spin-triplet component in an otherwise spin-singlet dominant junction, the current is immediately affected [red solid line of \fref{fig:current-phi}(b)]. The profile is still strongly non-sinusoidal, but smoothly turns more harmonic as the spin-triplet component becomes dominant. 
It is interesting to note that, even though the ABS in Eqs. (\ref{eq:ABS-cp1}) and (\ref{eq:ABS-cm1}) are $4\pi$-periodic, the dc current calculated in the thermodynamic equilibrium is $2\pi$-periodic. 
Within the thermodynamic equilibrium assumption, the occupation numbers of the subgap states remain fixed at a temperature $T$. As a consequence, the Josephson current depends on the temperature as shown in \eref{eq:current-abs-D1} and the factors $\tanh[E_i/(2k_BT)]$ directly affect the periodicity of the current, reducing it from $4\pi$ to $2\pi$. 

For junctions with arbitrary barrier strength $Z$, the contribution to the total current from the intermediate region $I_2(\phi)$ becomes, in general, non-zero. The contribution from the continuum $I_3(\phi)$, however, is zero for short ballistic junctions. The Josephson current is thus given by $I(\phi)=I_1(\phi)+I_2(\phi)$, including both spin channels. To compute the current, we use a general expression based on quasi-classical Green functions\cite{Furusaki_1991,Tanaka_1997,Asano_2001} (see details in Appendix \ref{sec:app2}). 
In \fref{fig:current-phi}(c,d) we repeat the previous cases for $Z=4$. 
In the presence of a barrier, the current profile becomes more harmonic due to the contribution from the intermediate region. 
Additionally, the chiral behavior of the spin-triplet component of the pair potential becomes more important when $\Delta_s<\Delta_p$. As a result, the amplitude of the current is greatly increased in this regime, for both chiralities [black dot-dashed lines in \fref{fig:current-phi}(c,d)]. 
Juntions with $\chi=+1$ feature the highest increase in the amplitude and almost harmonic profile. Junctions with $\chi=-1$, however, display a non-sinusoidal behavior in the $\Delta_s<\Delta_p$ regime. 


\begin{figure}
	\includegraphics[width=\columnwidth]{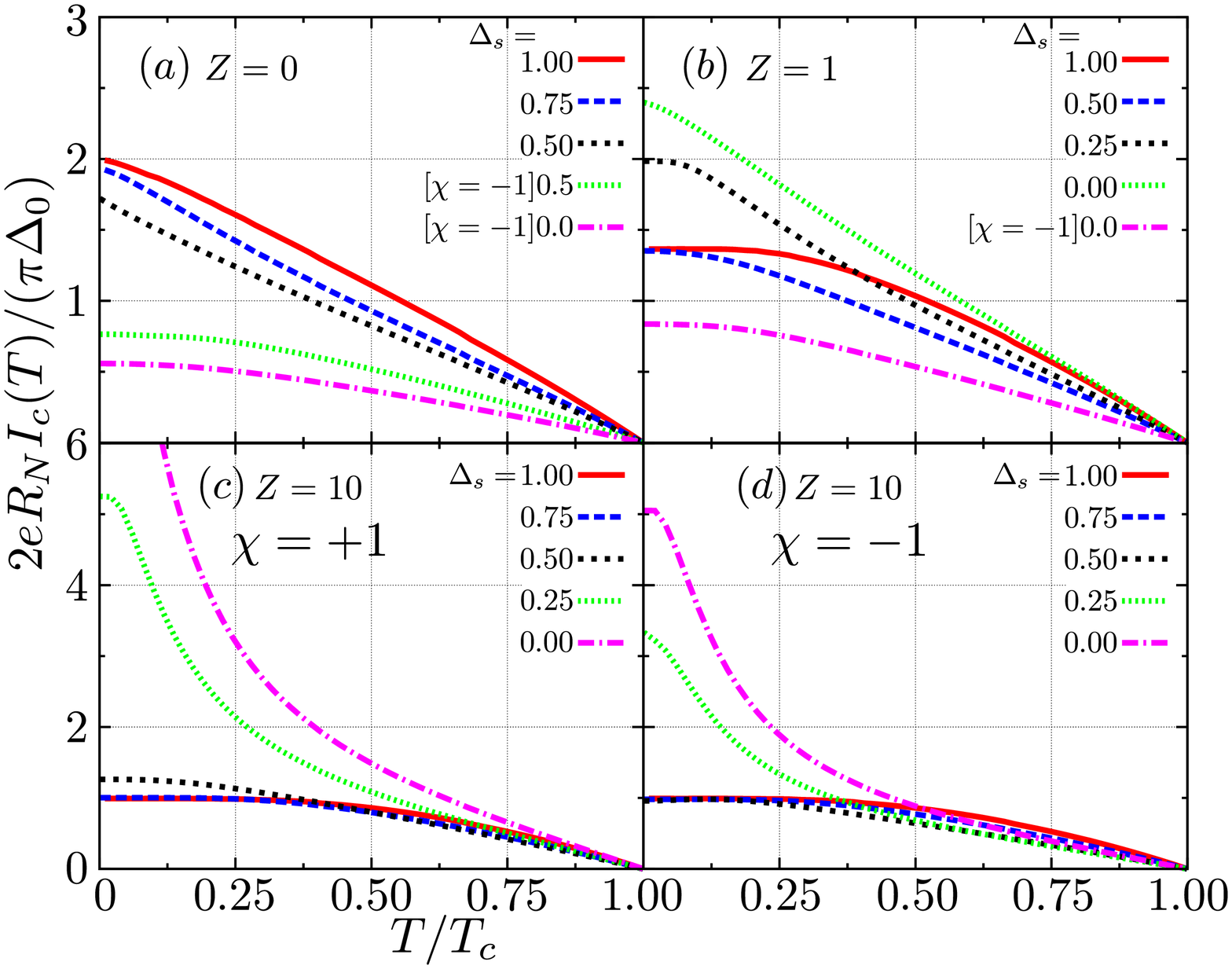}
	\caption{\label{fig:temp-Z}
	Maximum Josephson current $I_c(\phi)$ as a function of the temperature for different barrier strengths: $Z=0$ (a), $Z=1$ (b), and $Z=10$ (c) and (d). For (a) and (b), the junction chirality is $\chi=+1$ unless otherwise specified. In all cases the mixing between spin-singlet and spin-triplet is controlled by $\Delta_s=1-\Delta_p$.  Temperature is normalized to the critical temperature of $s$-wave superconductors $T_c=8.8K$. 
	}
\end{figure}


To study the effect of temperature on the Josephson current, we assume that the pair potentials $\Delta_{1,2}(T)$ have the standard BCS dependence. For simplicity, we only consider symmetric junctions. 
We show in \fref{fig:temp-Z} the dependence of the critical Josephson current $I_c(\phi)$ on the temperature for junctions with different barrier strengths. 
Transparent junctions between $s$-wave superconductors are equivalent to those between $p$-wave superconductors, when the angular momentum of the Cooper pairs align\cite{Yakovenko_2004,Asano_2002}. 
In this case, the effect of the mixing is to reduce the amplitude until a minimum is reached for $\Delta_s=\Delta_p=0.5$ [see \fref{fig:temp-Z}(a)]. For the opposite chirality, a minimum is reached for $\Delta_s=0$ and the effect of the spin-mixing is to smoothly reduce the critical current from $\Delta_s=1$. 

In \fref{fig:temp-Z}(b) we analyze junctions with a finite barrier ($Z=1$). 
The effect of the barrier is to saturate the critical current to a fixed value at low temperatures. The saturation point depends on the barrier strength, the mixing of spin states, and the relative chirality of the junction. 
Junctions with opposite chirality ($\chi=-1$) display a behavior similar to the perfectly transparent case, with a smooth transition between a maximum critical current for $\Delta_s=1$ and a minimum at $\Delta_s=0$. 
For $\chi=+1$, the minimum critical current is still found for $\Delta_s=0.5$. However, for $Z\neq0$, the critical current is enhanced in the spin-triplet dominant range $\Delta_s<\Delta_p$, compared to the $s$-wave behavior of $\Delta_s=1$. The current amplitude is maximum for $\Delta_s=0$. 

This tendency is greatly enhanced for tunnel junctions. In \fref{fig:temp-Z}(c,d) we go into this limit setting $Z=10$. We show the temperature dependence for junctions with $\chi=+1$ in \fref{fig:temp-Z}(c) and with $\chi=-1$ in \fref{fig:temp-Z}(d). 
For junctions with $\chi=+1$, the amplitude of the Josephson current is enhanced by the mixture of spin-states. The critical current becomes much larger than that of $s$-wave superconductors the moment the spin-triplet component becomes dominant ($\Delta_p>\Delta_s$). This enhancement of the critical current is found for a wide range of temperatures below the critical temperature $T_c$. 
A similar behavior is found at low temperatures for junctions with $\chi=-1$. In this case, however, the amplitude of the Josephson current is reduced with respect to that of $s$-wave superconductors when the temperature is comparable to $T_c$. 
At low temperatures, the impact of the mixing of spin-states is also different from that of $\chi=+1$ junctions. Namely, when a small spin-triplet component is added to both superconductors such that their chiralities are antiparallel, the critical current is immediately reduced. $I_c$ reaches a minimum when $\Delta_s=\Delta_p$, i.e., when the amplitude of both spin-states is the same. In the regime $\Delta_s<\Delta_p$, the critical current is greatly enhanced. 
The current amplitude for $\chi=+1$ is always bigger than the $\chi=-1$ case, independently of the mixture of spin-states. 
Following the study of Ref.~\onlinecite{Asano_2002} for pure $p$-wave junctions ($\Delta_s=0$), this is a direct consequence of the zero-energy states formed at the junction. For $\chi=+1$, these states are formed from constructive interference of quasiparticles from both superconductors [see \fref{fig:ABS-gap}(a)]. On the other hand, quasiparticles interfere destructively when $\chi=-1$. 

\section{Conclusions\label{sec:conc}}
We have analyzed transport signatures of NS and Josephson junctions where the superconducting pairing potential shows a mixture of singlet and chiral triplet spin-states. For the spin-triplet part, we have studied an out-of-plane polarization where the pairing only affects the $\up\dw$ and $\dw\up$ spin channels. In this situation, the BdG equations are decoupled for these channels. 
For both spin channels, excitations are affected by one of two effective gaps $|\Delta_{1,2}(\theta)|=|\Delta_s\pm\Delta_p\e^{i\theta}|$ depending on their direction of motion. Additionally, the bulk gap of the superconductor is given by the smallest of these gaps, $|\Delta_{2}(\theta)|=|\Delta_s-\Delta_p\e^{i\theta}|$, and can be zero when $\Delta_s=\Delta_p$. As a consequence, two topologically different regions are defined: a trivial region with $\Delta_s>\Delta_p$ and a non-trivial one for $\Delta_s<\Delta_p$. 
Transport properties depend on which topological region the pair potential is in. 
For the trivial case with $\Delta_s>\Delta_p$, the NS conductance features a gap, while a zero-bias peak appears when $\Delta_s<\Delta_p$. In the latter case, the pair potential becomes complex and allows to form sub-gap resonances that contribute to the zero-bias peak. 
The double gap structure resulting from the mixing of spin-states can be detected in NS spectroscopy measurements. 

The formation of ABS at short ballistic SNS junctions is also affected by mixing of spin-states: in the topologically non-trivial regime with $\Delta_s<\Delta_p$ the ABS develop zero-energy states. 
The relative chirality of the superconductors at each side of the junction also plays an important role. 
For junctions where the angular momentum of the Cooper pairs align in parallel (i.e., $\chi=+1$), the inclusion of mixing of spin-states increases the amplitude of the critical current for any value of the mixing. 
When they align in opposite directions ($\chi=-1$), the critical current is only increased in the regime $\Delta_s<\Delta_p$. 
The zero-energy modes are also affected by the chirality of the junction. For $\chi=+1$, the zero-energy ABS only appear at $\phi=\pi$ for any incidence angle $\theta$ if $Z=0$ or for $\theta=0$ when $Z\neq0$. 
When $\chi=-1$, the zero-energy ABS appear for a wide range of angles of incidence, independently of the barrier strength $Z$. 

An out-of-plane polarization for the Cooper pairs is the only possible triplet state that decouples the BdG equations into two independent spin channels while still considering a mixed singlet-triplet pairing term. 
This restriction has allowed us to make analytical predictions and to better understand the underlying physics. 
A system that is considered to present a similar mixture of singlet and chiral triplet pairing is the eutectic phase of the Sr$_2$RuO$_4$-Ru. Sr$_2$RuO$_4$ is widely believed to be a chiral $p$-wave superconductor\cite{Mackenzie_2003,Maeno_2012}. The increased critical temperature for the Sr$_2$RuO$_4$-Ru eutectic system, however, is assumed to come from the interplay between the chiral $p$-wave order parameter of Sr$_2$RuO$_4$ and the conventional $s$-wave one of Ru\cite{Kaneyasu_2010,*Kaneyasu_2010b,Maeno_2012}. 
In the future, we would like to analyze the robustness of our results under the choice of more complicated (but maybe also more realistic) mixed pairing terms. 


\section*{Acknowledgments}
We thank Y. Asano, J. C. Budich, and G. Tkachov for helpful discussions and comments. 
We acknowledge financial support by the German-Japanese research unit FOR1483 on ``Topotronics''.


\appendix

\section{Josephson current from scattering theory\label{sec:app1}}
In this appendix, we calculate the Josephson current using the electrical current density, following Refs.~\onlinecite{BTK,vanWees_1991,Bagwell_1992} (see also Ref.~\onlinecite{Chang_1994} for an asymmetric junction). We demonstrate that the contribution to the \textit{total} Josephson current for $|E|>|\Delta_{2}(\theta)|$ is zero in a short ballistic junction with $Z=0$ (perfect transparency). The Josephson current also vanishes for junctions with arbitrary barrier strength $Z$ in the continuum region with $|E|>|\Delta_{1}(\theta)|$. 
For each spin channel separately, however, the current may be finite in both the intermediate and the continuum energy regions. 

The Josephson junction is modeled as in the main text with $\Delta_{s,p}^{\sigma,L}=\Delta_{s,p}^{\sigma,R}$ (symmetric junction). We consider four scattering processes: 
\begin{inparaenum}[(\itshape 1\upshape)]
\item an electron-like excitation incoming from the left superconductor; 
\item a hole-like excitation incoming from the left superconductor; 
\item ; and
\item the same processes with incidence from the right superconductor.
\end{inparaenum} 
In the first situation, the wave function from the left superconductor includes the incident electron-like excitation from spin channel $\sigma=1,2$ and the Andreev reflected one
\begin{align*}
 \Psi^{L}_{\sigma}(x) ={}& 
\left(\!\! \begin{array}{c}
          u_{\sigma}(\theta_{+}) \e^{i\phi} \\ \eta^{*}_{\sigma}(\theta_{+})v_{\sigma}(\theta_{+})
\end{array}\!\! \right)  \e^{ikx}\e^{-\frac{\Omega_{\sigma}(\theta_{+})x}{\hbar v_F}} \\ {}& + a^{(1)}_{\sigma}\left(\!\! \begin{array}{c}
          \eta_{\sigma}(\theta_{+})v_{\sigma}(\theta_{+}) \\ u_{\sigma}(\theta_{+}) 
\end{array}\!\! \right)  \e^{ikx} \e^{\frac{\Omega_{\sigma}(\theta_{+})x}{\hbar v_F}} \,\, . 
\end{align*}
The transmitted excitation in the right superconductor is
\begin{equation*}
  \Psi^{R}_{\sigma} = 
c^{(1)}_{\sigma} \left(\!\! \begin{array}{c}
          u_{\sigma}(\theta_{+}) \e^{i\phi} \\ \eta^{*}_{\tilde{\sigma}}(\theta_{+})v_{\sigma}(\theta_{+})
\end{array}\!\! \right)  \e^{ikx}\e^{-\frac{\Omega_{\sigma}(\theta_{+})x}{\hbar v_F}} \,\,\, .
\end{equation*}

As in the main text, we are using the Andreev approximation so the wave vector $k$ is the same for both electrons and holes. The wave functions for the remaining processes are analogously obtained from the solutions of the BdG equations given in \eref{eq:wave-functions_J}. 
Enforcing continuity of the wave functions at $x=0$ for each case, we obtain
\begin{align*}
 c^{(1)}_{\sigma}=\e^{-i\phi}c^{(4)}_{\sigma}=\frac{u_{\sigma}^2(\theta) - v_{\sigma}^2(\theta)}{\e^{i\phi}u_{\sigma}^2(\theta) - v_{\sigma}^2(\theta)} \quad , \\
 c^{(3)}_{\sigma}=\e^{i\phi}c^{(2)}_{\sigma}=\frac{u_{\bar{\sigma}}^2(\theta) - v_{\bar{\sigma}}^2(\theta)}{\e^{-i\phi}u_{\bar{\sigma}}^2(\theta) - v_{\bar{\sigma}}^2(\theta)} \quad , 
\end{align*} 
with $\bar{\sigma}=1,2\neq\sigma$. Therefore, for the perfectly transparent case we have $|c^{(1)}_{\sigma}|^2=|c^{(4)}_{\sigma}|^2=|c^{(2)}_{\bar{\sigma}}|^2=|c^{(3)}_{\bar{\sigma}}|^2$. 

We can now define the electrical current transmission amplitude for an electron-like excitation incident from the left superconductor with $|E|>|\Delta_{2}|$ as 
\begin{equation*}
 T_{L\rightarrow R}^{e\sigma}=\frac{1}{2}\left[ \left(|u_{\sigma}(\theta)|^2 + |v_{\sigma}(\theta)|^2 \right)f(E) - |v_{\sigma}(\theta)|^2 \right] |c^{(1)}_{\sigma}|^2 \,\, ,
\end{equation*}
with $f(E)=[1+\exp(E/k_BT)]^{-1}$ the Fermi occupation factor, $T$ the absolute temperature and $k_B$ the Boltzmann constant. 
For the rest of the processes, we find
\begin{align*}
  T_{L\rightarrow R}^{h\sigma}={}&\frac{1}{2}\left[ \left(|u_{\bar{\sigma}}(\theta)|^2 + |v_{\bar{\sigma}}(\theta)|^2 \right)f(E) - |u_{\bar{\sigma}}(\theta)|^2 \right] |c^{(1)}_{\bar{\sigma}}|^2 \,\, , \\
  T_{R\rightarrow L}^{e\sigma}={}&\frac{1}{2}\left[ \left(|u_{\bar{\sigma}}(\theta)|^2 + |v_{\bar{\sigma}}(\theta)|^2 \right)f(E) - |v_{\bar{\sigma}}(\theta)|^2 \right] |c^{(1)}_{\bar{\sigma}}|^2 \,\, , \\
  T_{R\rightarrow L}^{h\sigma}={}&\frac{1}{2}\left[ \left(|u_{\sigma}(\theta)|^2 + |v_{\sigma}(\theta)|^2 \right)f(E) - |u_{\sigma}(\theta)|^2 \right] |c^{(1)}_{\sigma}|^2 \,\, .
\end{align*}

The electrical current operator associated to process {\it (1)} is defined as $J_{\sigma}^{(1)}=(e\hbar/m)k_{\sigma}^{(1)}T_{L\rightarrow R}^{e\sigma}$, with $k_{\sigma}^{(1)}=+k$ under Andreev approximation. Analogously, we find $k_{\sigma}^{(4)}=+k$ and $k_{\sigma}^{(2)}=k_{\sigma}^{(3)}=-k$. 
Finally, the electrical current per unit energy carried by electron and hole-like quasiparticles is given by
\begin{align*}\label{eq:current}
 I^{e,h}_{\sigma}(\phi)={}&\frac{e}{2h} \int^{\infty}_{|\Delta_2(\theta)|} \left[ \rho^{L}_{\sigma}(E,\theta) T_{L\rightarrow R}^{(e,h)\sigma}(E,\phi) \right. \nonumber \\
 {}& \left. - \rho^{R}_{\sigma}(E,\theta) T_{R\rightarrow L}^{(e,h)\sigma}(E,\phi) \right]\mathrm{d}E  \,\, ,
\end{align*}
with 
\[
 \rho^{L,R}_{\sigma}(E,\theta)=\lim\limits_{\eta\rightarrow0}\imag\left\{\frac{E+i\eta}{\sqrt{|\Delta^{L,R}_{\sigma}(\theta)|^2-(E+i\eta)^2}} \right\}
\]
the normalized density of states in the superconductor. 

For each spin channel, we define the total electrical current per unit energy as $I_{\sigma}(\phi)=[I^{e}_{\sigma}(\phi)+I^{h}_{\sigma}(\phi)]/2$. For perfect transparent junctions, we find that $I_{\sigma}(\phi)=-I_{\bar{\sigma}}(-\phi)$. As a consequence, the contribution to the total Josephson current for $|E|>|\Delta_2(\theta)|$ vanishes, although the current from each spin channel can be finite. This result is also valid at arbitrary junction transparency for the regime $|E|>|\Delta_1(\theta)|$ (continuum). 
For a perfectly transparent junction, therefore, the Josephson current is given only by the contribution from the Andreev bound states. For junctions with $Z\neq0$, we must also include the contribution from the intermediate region. 
On the other hand, for each spin channel the current must include the terms from the three energy regions. 

\section{Josephson current from quasi-classical Green functions\label{sec:app2}}
In this Appendix we provide the main details to adapt the formula for the dc Josephson current derived in Refs.~\onlinecite{Furusaki_1991,Tanaka_1997,Asano_2001} to the present work. For a junction between superconductors with mixed singlet and triplet spin states as described in the main text, a compact expression for the Josephson current can be taken from Ref.~\onlinecite{Tanaka_1997} as
\begin{gather}
 R_N I(\phi)= \nonumber \\ \frac{\pi\bar{R}_N k_B T}{e} \left\{ \sum\limits_{\omega_n} \int_{-\pi/2}^{\pi/2} \bar{F}(\theta,i\omega_n,\phi) \cos\theta \mathrm{d}\theta \right\} \,\, 
\label{eq:current-TK}
\end{gather}
with
\begin{equation*}
  \bar{R}_N = \left(\int_{-\pi/2}^{\pi/2} \sigma_N \cos\theta \mathrm{d}\theta \right)^{-1} \,\, ,
\end{equation*}
and
\begin{equation*}
 \sigma_N= \frac{4\cos^2\theta}{4\cos^2\theta + Z^2} \quad . 
\end{equation*}
The integrand is given by
\begin{gather*}
 \bar{F}(\theta,E,\phi)= \\ \sum\limits_{\sigma}  \left\{ |\Delta^L_{\sigma}(\theta_+)|\frac{a^{(1)}_{\sigma}(\theta,E,\phi)}{\Omega^{L}_{\sigma}(\theta_{+})} 
 - |\Delta^L_{\sigma}(\theta_-)|\frac{a^{(2)}_{\sigma}(\theta,E,\phi)}{\Omega^{L}_{\sigma}(\theta_{-})} \right\} \, .
\end{gather*}
The Andreev reflection amplitudes $a^{(1,2)}_{\sigma}(\theta,E,\phi)$ are obtained after substituting the wave functions from \eref{eq:wave-functions_J} into the boundary conditions of \eref{eq:bbcc}. In order to use them in \eref{eq:current-TK}, we apply the analytical continuation $E\rightarrow i\omega_n$, where $\omega_n=\pi k_B T (2n+1)$ denotes the Matsubara frequency. Subsequently, we write $\Omega^{L,R}_{\sigma}(\theta_{\pm})=\sgn(\omega_n)\sqrt{|\Delta^{L,R}_{\sigma}(\theta_{\pm})|^2+w_n^2}$. Finally, for the pairing potentials, we use 
\begin{align*}
 \Delta^L_{\up\dw}(\theta_{\pm})={}& \Delta^L_s \pm \Delta^L_p\e^{\pm i\theta} \quad , \\
 \Delta^L_{\dw\up}(\theta_{\pm})={}& -\left(\Delta^L_s \mp \Delta^L_p\e^{\pm i\theta} \right) \quad , \\
 \Delta^R_{\up\dw}(\theta_{\pm})={}& \Delta^R_s \pm \Delta^R_p\e^{\pm i\chi\theta} \quad , \\
 \Delta^R_{\dw\up}(\theta_{\pm})={}& -\left(\Delta^R_s \mp \Delta^R_p\e^{\pm i\chi\theta} \right) \quad .
\end{align*}


%


\end{document}